\documentclass[aps,prl,floatfix,showpacs,twocolumn,preprintnumbers,tightenlines,superscriptaddress]{revtex4-1}

\RequirePackage{epsfig}
\RequirePackage{graphicx}

\RequirePackage[centertags]{amsmath}
\RequirePackage{amssymb}
\RequirePackage{color}
\RequirePackage{siunitx}

\RequirePackage{tabularx}
\RequirePackage{rotating}
\RequirePackage{multirow}
\RequirePackage{tabu}
\RequirePackage{hhline}

\RequirePackage{enumerate}
\RequirePackage{enumitem}

\RequirePackage[normalem]{ulem}
\RequirePackage{wrapfig}
\RequirePackage{verbatim}

\RequirePackage[colorlinks=true,linkcolor=blue,urlcolor=blue,hyperfootnotes=false,citecolor=black]{hyperref}
\RequirePackage{hypernat} 
\allowdisplaybreaks[1]
\begin{document}
\title{Tuneable hopping and nonlinear cross-Kerr interactions in a high-coherence superconducting circuit}

\author{M.~Kounalakis} 
 \email{marios.kounalakis@gmail.com}
\affiliation{Kavli Institute of Nanoscience, Delft University of Technology, Lorentzweg 1, 2628 CJ Delft, The Netherlands}
 \affiliation{QuTech, Delft University of Technology, Delft, The Netherlands}
\author{C.~Dickel}
\affiliation{QuTech, Delft University of Technology, Delft, The Netherlands}
\affiliation{Kavli Institute of Nanoscience, Delft University of Technology, Lorentzweg 1, 2628 CJ Delft, The Netherlands}
\author{A.~Bruno}
\affiliation{QuTech, Delft University of Technology, Delft, The Netherlands}
\affiliation{Kavli Institute of Nanoscience, Delft University of Technology, Lorentzweg 1, 2628 CJ Delft, The Netherlands}
\author{N.~K.~Langford}
\affiliation{QuTech, Delft University of Technology, Delft, The Netherlands}
\affiliation{Kavli Institute of Nanoscience, Delft University of Technology, Lorentzweg 1, 2628 CJ Delft, The Netherlands}
\affiliation{School of Mathematical and Physical Sciences, University of Technology Sydney, Ultimo, New South Wales 2007, Australia}
\author{G.~A.~Steele}
\affiliation{Kavli Institute of Nanoscience, Delft University of Technology, Lorentzweg 1, 2628 CJ Delft, The Netherlands}

\begin{abstract}
Analog quantum simulations offer rich opportunities for exploring complex quantum systems and phenomena through the use of specially engineered, well-controlled quantum systems.
A critical element, increasing the scope and flexibility of such experimental platforms, is the ability to access and tune \textit{in situ} different interaction regimes.
Here, we present a superconducting circuit building block of two highly coherent transmons featuring \textit{in situ} tuneable photon hopping and nonlinear cross-Kerr couplings.
The interactions are mediated via a nonlinear coupler, consisting of a large capacitor in parallel with a tuneable superconducting quantum interference device (SQUID).
We demonstrate the working principle by experimentally characterising the system in the single- and two-excitation manifolds, and derive a full theoretical model that accurately describes our measurements.
Both qubits have high coherence properties, with typical relaxation times in the range of 15 to 40 microseconds at all bias points of the coupler.
Our device could be used as a scalable building block in analog quantum simulators of extended Bose-Hubbard and Heisenberg XXZ models, and may also have applications in quantum computing such as realising fast two-qubit gates and perfect state transfer protocols.
\end{abstract}

\maketitle

\section{Introduction} 

Analog quantum simulations, where engineered systems emulate the behaviour of other, less accessible quantum systems in a controllable and measurable way~\cite{feynman1982simulating}, show significant promise for improving our understanding of complex quantum phenomena without the need for a full fault-tolerant quantum computer~\cite{broome2013photonic, spring2013boson, bernien2017probing, zhang2017observation}.
In this paradigm, the versatility of the simulator is determined by the range of interaction types and complexity accessible to the emulating quantum system.
Promising avenues for pushing beyond what can be simulated with a classical machine include the study of highly interacting many-body systems~\cite{cirac2012goals, houck2012on-chip, georgescu2014quantum, noh2016quantum, hartmann2016quantum}.
Superconducting circuit quantum electrodynamics (QED) is a very attractive platform for analog quantum simulation because of site-specific control and readout, and because of the flexible and engineerable system designs, which have led to the study of many interesting effects~\cite{li2013motional, chen2014emulating, salathe2015digital, eichler2015exploring, roushan2017chiral, langford2017experimentally, braumuller2017analog, fitzpatrick2017observation, roushan2017spectral, xu2018emulating, potocnik2018studying, kandala2017hardware-efficient}.
Adding new components to the circuit QED design toolbox such as novel types of interactions can dramatically increase the range of phenomena that can be simulated~\cite{jin2013photon, jin2014steady-state, marcos2014two}.
For example, for exploring exotic effects, like quantum phase transitions in systems of strongly correlated particles, it is important to be able to access and rapidly tune between different many-body interaction regimes.

\textit{In situ} tuneable couplers have been successfully realised in a variety of circuit QED architectures~\cite{bertet2006parametric, hime2006solid-state, harris2007sign, niskanen2007quantum, vanderPloeg2007controllable, allman2010rf-SQUID, bialczak2011fast, srinivasan2011tunable, wulschner2016tunable, weber2017coherent}, in particular using the more coherent transmon design~\cite{koch2007charge, chen2014qubit, mckay2016universal, lu2017universal}.
In recent experiments, transmon arrays with tuneable exchange-type hopping interactions, have been employed to study many-body localisation phenomena of Bose-Hubbard and spin-1/2 XY models~\cite{roushan2017spectral, xu2018emulating}.
However, moving beyond linear couplings to incorporate additional nonlinear interactions would enable the emulation of far more complex Hamiltonians.
For example, nonlocal cross-Kerr interactions, present in extended Bose-Hubbard models~\cite{vanOtterlo1995quantum, mazzarella2006extended}, introduce much richer many-body phase diagrams, leading to intriguing phenomena such as crystalline and supersolid phases of light as the ratio of the hopping and cross-Kerr coupling strengths is varied~\cite{jin2013photon, jin2014steady-state}.
In the qubit context, nonlinear cross-Kerr coupling, sometimes referred to as longitudinal coupling, is essential for engineering plaquette interactions in lattice gauge theories~\cite{marcos2014two} and gives access to a large class of quantum-dimer and XYZ spin-model Hamiltonians.

Here, we demonstrate tuneable hopping and cross-Kerr interactions in a highly coherent two-transmon unit cell.
Specifically, using a large capacitor in parallel with a tuneable nonlinear inductor as a coupling element, we are able to tune the ratio of the two coupling strengths, even suppressing hopping completely while maintaining a non-zero cross-Kerr coupling, giving access to different interaction regimes.
We comprehensively characterise the energy landscape of this building block using different spectroscopic techniques.
We show excellent agreement with a full theoretical model we have developed to describe the underlying circuit Hamiltonian including higher transmon excitation manifolds.
Finally, we have thoroughly studied the qubit coherence as a function of the coupler bias, showing high relaxation times of $15-40~\mu$s, and dephasing times reaching up to $40~\mu$s at flux-insensitive operating points.
Our work outlines a new recipe for building scalable analog quantum simulators of complex Hamiltonians using coupled transmon arrays, and our theoretical model provides an invaluable tool for designing and realising larger scale implementations.

 \begin{figure}[t]
  \centering
  \includegraphics[width=\linewidth]{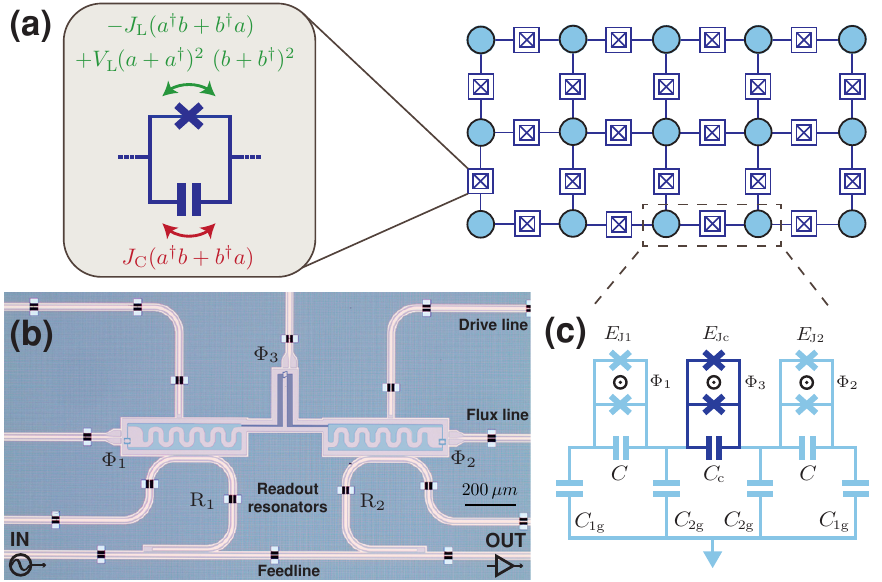}
  \caption{
  {\bf Working principle and experimental device.}
{\bf (a)}~A nonlinear coupler introducing hopping and cross-Kerr interactions between transmons (circles) on a circuit QED lattice.
Photon hopping, mediated by the capacitor and the inductor (Josephson junction), can be coherently suppressed at the filtering condition ($J_\text{L}=J_\text{C}$), in analogy with the working principle of an LC filter.
Tuning the nonlinear inductance can enable interesting regimes where cross-Kerr and photon-pair tunnelling dynamics are equivalent or even dominant over photon hopping processes.
{\bf (b)}~Optical micrograph of the experimental device with added false-colour on the transmon-coupler superconducting islands.
Qubit readout and microwave control is performed via dedicated resonators $R_{1,2}$ that are coupled to a common feedline.
Dedicated drive lines provide additional microwave control to each transmon.
On-chip flux bias lines are used to tune the qubit frequencies and their mutual coupling.
The scale bar corresponds to $200~\mu$m.
{\bf (c)}~Circuit diagram of the implemented building block.
The coupler (dark blue) is realised using a capacitor $C_\text{c}$ in parallel with a tuneable nonlinear inductor (SQUID) that is galvanically connected with the two transmon qubits (light blue).}
  \label{fig:Chip}
\end{figure}

\section{Results}

\subsection{Implementing nonlinear couplings}

The working principle of the coupler is similar to that of a band-stop LC filter, relying on the fact that its impedance $Z(\omega)=\frac{-i\omega}{C(\omega^2-\omega_{\text{LC}}^2)}$ is infinite on resonance, as currents through the capacitor and the inductor interfere destructively.
We implement a nonlinear analog of this circuit by using a nonlinear superconducting quantum interference device (SQUID) as the inductor, realising tuneable cross-Kerr and nearest-neighbour hopping interactions (Fig.~\ref{fig:Chip}a).
The Josephson nonlinearity of the SQUID gives rise to tuneable higher-order nonlocal terms~\cite{vanOtterlo1995quantum}, including cross-Kerr interactions, which are equivalent to longitudinal $\hat{\sigma}_z\hat{\sigma}_z$ coupling in the qubit subspace.
By contrast, the linear single-excitation hopping between the two sites is mediated by both the capacitor $C_\text{c}$ at a constant rate $J_\text{C}$, and by the inductor at a tuneable rate $-J_\text{L}$.
Because of interference between these two processes, the hopping strength tunes in a different way from the cross-Kerr coupling, making different many-body interaction regimes accessible.
In particular, at the point where the hopping rates cancel ($J_\text{C}=J_\text{L}$), we can access a purely nonlinear regime with zero linear interaction.

The nonlinear coupler is implemented in a circuit QED device of two superconducting transmon qubits, the basic building block required for future lattice implementations (Fig.~\ref{fig:Chip}a).
The optical micrograph along with a circuit diagram of the device are shown in Figs.~\ref{fig:Chip}b and \ref{fig:Chip}c, respectively.
Each transmon, consisting of two superconducting islands connected by an interdigitated capacitor $C$ and a tuneable SQUID inductance, resonates at a plasma frequency $\omega\simeq(\sqrt{8E_{\text{J}}E_{\text{C}}}-E_{\text{C}})/\hbar$, with Josephson energy $E_{\text{J}}$ and charging energy $E_{\text{C}}=\frac{e^2 }{2 \widetilde{C}}$, where $\widetilde{C}$ is the effective transmon capacitance (see Supplementary Eq.~(S12)).
Transmon frequencies can be independently tuned using on-chip flux lines ($\Phi_{1,2}$), and spectroscopy is performed through dispersively coupled readout resonators ($R_{1,2}$) measured via a common microwave feedline (see Supplementary~Fig.~S5 for the full measurement setup and Fig.~S6 for qubit spectroscopy vs $\Phi_{1,2}$).
Microwave drives are applied via either the resonators or dedicated drivelines.
The coupler capacitance $C_\text{c}$ and flux-tuneable SQUID ($\Phi_{3}$) connect galvanically to the two transmons.

The physics of the two-transmon building block is, to a good approximation, well described by an extended Bose-Hubbard model, which is a Heisenberg XXZ spin model in the qubit regime.
To achieve this, we need to operate the qubits detuned from coupler resonances, so that coupler excitations do not participate directly in the system dynamics.
Under this condition, the system can be described by a simplified two-transmon Hamiltonian:

\begin{align}
\label{eq:H}
\begin{split}
\hat{\mathcal{H}}/\hbar ~=~&\omega_{\rm 1}~\hat{a}^\dagger \hat{a}-U~\hat{a}^\dagger \hat{a}^\dagger \hat{a}\hat{a}\\
&+\omega_{\rm 2}~\hat{b}^\dagger \hat{b}-U~\hat{b}^\dagger \hat{b}^\dagger \hat{b}\hat{b}\\
&+ J(\hat{a}^\dagger \hat{b}+\hat{b}^\dagger \hat{a})\\
&+ V ~\hat{a}^\dagger \hat{a} \hat{b}^\dagger \hat{b},
\end{split}
\end{align}
where $\hat{a}^{(\dagger)},~\hat{b}^{(\dagger)}$ are bosonic annihilation (creation) operators for each transmon in the uncoupled basis, with on-site nonlinearity $U=\frac{E_{\text{C}}}{2 \hbar}$.
The interaction between the two sites can be described by hopping of single excitations at a rate 
\begin{align}
\label{eq:transverse}
J~=~J_\text{C}-J_\text{L}~=~\frac{\sqrt{8{E}_{\text{J}}E_{\text{C}}}}{2\hbar}\left(\frac{C_\text{c}}{4C_{\text{eff}}}-\frac{E_{\text{J}}^{\text{c}}}{4{E}_{\text{J}}}\right),
\end{align}
and a cross-Kerr coupling strength
\begin{align}
\label{eq:cross-Kerr}
V=  -\frac{E_{\text{J}}^{\text{c}}E_\text{C}}{8\hbar{E}_{\text{J}}}.
\end{align}
The capacitive coupling $J_\text{C}$ is fixed and determined by the ratio of the coupling capacitor $C_\text{c}$ to an effective capacitance $C_{\text{eff}}$, which depends on the circuit network (see Supplementary Eq.~(S27)).
The interaction strengths $J_\text{L}$ and $V$ are determined by tuning the Josephson energy of the coupling SQUID, $E_{\text{J}}^{\text{c}}=E_{\text{J}}^{\text{c~(max)}}\cos{(\pi\Phi_3/\Phi_0)}$.

Importantly, the cross-Kerr coupling $V$ is different from the diagonal coupling that can be observed in linearly coupled transmon architectures, where the self-Kerr nonlinearity of each transmon leads to an effective cross-Kerr coupling between the normal modes of the system.
Such effective diagonal coupling scales with the hopping interaction strength and vanishes at $J=0$~\cite{geller2015tunable}, making it impossible to tune the ratio $J/V$ independently.
In our design, however, the cross-Kerr interaction results directly from the nonlinearity of the coupling junction and tunes to zero at a different coupler bias from the linear hopping interaction, giving access to different interaction regimes.
As $\Phi_3$ is tuned towards the filtering condition ($J_\text{L}\sim J_\text{C}$), the linear hopping term is suppressed more rapidly than the nonlinear cross-Kerr coupling, allowing access to the $J \ll V$ regime.
By contrast, when $\Phi_3 \sim 0.5\,\Phi_0$, the cross-Kerr coupling is suppressed ($V=J_\text{L}=0$) and the dynamics are dominated by single-photon hopping at a rate $J_\text{C}$.

In the full treatment of the quantum circuit, the nonlinear inductance also gives rise to correlated hopping and two-photon tunnelling correction terms, which play a role in the higher-excitation manifold.
These terms may also lead to interesting physical phenomena, which we return to in the discussion section.
We derive the full quantum model in the Supplementary Information~\cite{TunCouplerSOM}, along with a classical normal mode analysis providing supporting intuition for the full system behaviour.

\subsection{Tuneable single-photon hopping} 

\begin{figure}[t]
  \begin{center}
  \includegraphics[width=1\linewidth]{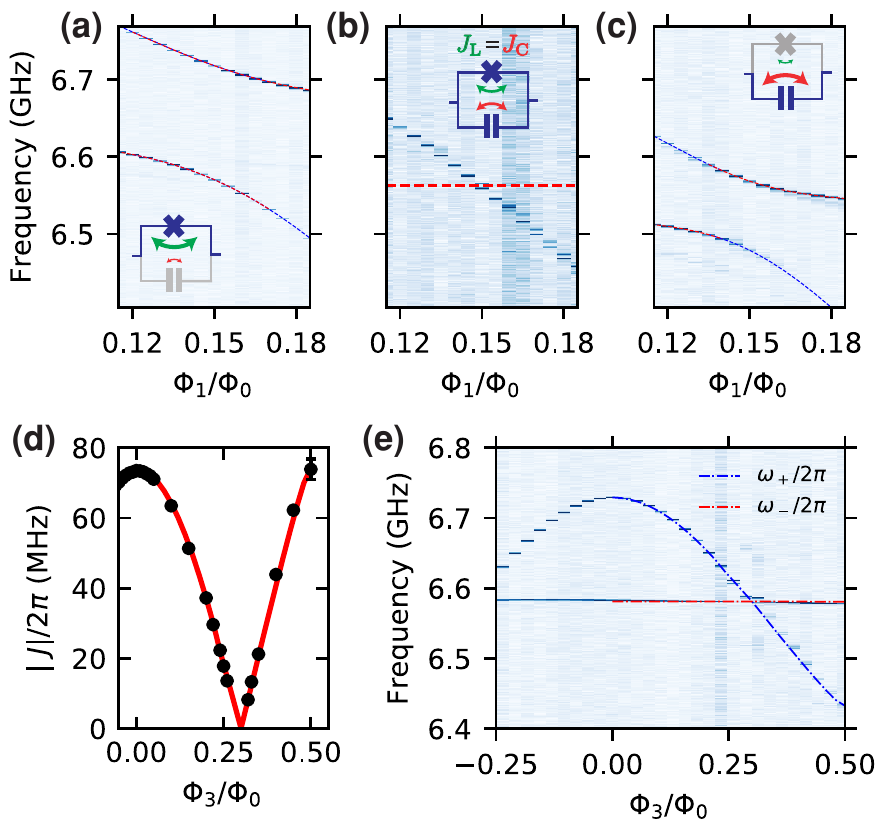}
  \end{center}
  \caption{
  {\bf Tuneable linear coupling and single-photon hopping suppression. }
Top: Avoided crossings for {\bf (a)}~inductive, {\bf (b)}~zero and {\bf (c)}~capacitive coupling regimes.
In all three cases, the frequency of qubit 2 is set around 6.6~GHz ($\Phi_2=0$), while we tune qubit 1 through resonance.
Qubit spectroscopy is performed via readout resonator $R_2$ in {\bf (a)} and {\bf (c)}, and via $R_1$ in {\bf (b)}.
The coupling elements that participate more strongly to the qubit-qubit interaction are indicated in the insets.
{\bf (d)}~Linear coupling strength $|J|$ obtained from a series of fitted avoided crossings, at different values of calibrated coupler flux bias $\Phi_3$.
{\bf (e)}~Eigenspectrum of the coupled qubit system on resonance vs $\Phi_3$, fitted with a simplified circuit theory model (the fitting parameters are listed in Table~\ref{tab:Parameters}).
The normal-mode splitting gets suppressed at the crossover between inductively and capacitively dominated coupler regimes $(\Phi_3/\Phi_0\simeq0.3)$.
Cross-talk effects between the different flux channels have been calibrated out experimentally.}  \label{fig:TunCoupling}
\end{figure}

We demonstrate our ability to tune the linear single-photon hopping interaction between the two transmons, by measuring qubit-qubit avoided crossings, in Fig.~\ref{fig:TunCoupling}.
The top panels show example crossings, as qubit 1 is tuned on resonance with the other at $\sim6.6$~GHz, in three different coupling scenarios, $J_\text{L}>J_\text{C}$ in Fig.~\ref{fig:TunCoupling}a, $J=0$ ($J_\text{L}=J_\text{C}$) in Fig.~\ref{fig:TunCoupling}b, and the $J_\text{C}$-dominated regime in Fig.~\ref{fig:TunCoupling}c.
The measurements in Figs.~\ref{fig:TunCoupling}a,~c are performed via readout resonator $R_2$, while in the zero coupling case (Fig.~\ref{fig:TunCoupling}b) we measure via $R_1$.
The range of typical coupling strengths that can be achieved with this device is illustrated in Fig.~\ref{fig:TunCoupling}d, where we plot $|J|/2\pi$ vs calibrated coupler bias $\Phi_3$.
We note that we have measured larger coupling strengths, up to 140~MHz, when operating at different qubit frequencies $\sim 5.4$~GHz (see Supplementary Fig.~S3).
The linear coupling is suppressed when the qubit frequencies are equal to the filter frequency $1/\sqrt{L_\text{c}C_\text{c}}$, which here takes place around $\Phi_3\sim0.3\ \Phi_0$.
Note that a higher transition of a lower-frequency \emph{sloshing} mode of the circuit, crossing with the qubits around this point, limits the observed on-off ratio in this device to $\sim 10$ (see Supplementary Fig.~S3).
This low-frequency mode, hybridising with the qubits, is associated with currents flowing only through the coupling inductor~\cite{TunCouplerSOM}, and could be avoided with slightly different design parameters (see Supplementary Fig.~S4).
Additional avoided crossing measurements in the regime where the linear coupling gets suppressed and reverses sign are plotted in Supplementary Fig.~S9, with spectroscopy on both qubits.

For a more complete characterisation of the tuneable hopping interaction, we fit the experimentally measured coupled-qubit spectrum with our theoretical model of the quantum circuit.
More specifically, in Fig.~\ref{fig:TunCoupling}e, we fix the qubits on resonance and plot the normal-mode splitting between the dressed states $|+\rangle~=~\frac{|01\rangle+|10\rangle}{\sqrt{2}}$ and $|-\rangle~=~\frac{|01\rangle-|10\rangle}{\sqrt{2}}$, as we tune the calibrated coupler flux bias.
The blue and red curves are theory fits of the single-excitation qubit manifold to the quantum circuit  Hamiltonian (Eq.~(S11) in the Supplementary), showing excellent agreement with the experimentally obtained spectrum.
The parameters obtained from this fit, which neglects higher-order couplings of each transmon to the low-frequency sloshing mode, are listed in Table~\ref{tab:Parameters}.
Note that the antisymmetric mode frequency $\omega_-/2\pi$ is unaffected by coupler tuning, which reflects the fact that this mode is only associated with charge oscillations across the qubit junctions (see Supplementary Fig.~S2).

\subsection{Tuneable nonlinear cross-Kerr coupling}

\begin{figure}[t]
  \begin{center}
  \includegraphics[width=\linewidth]{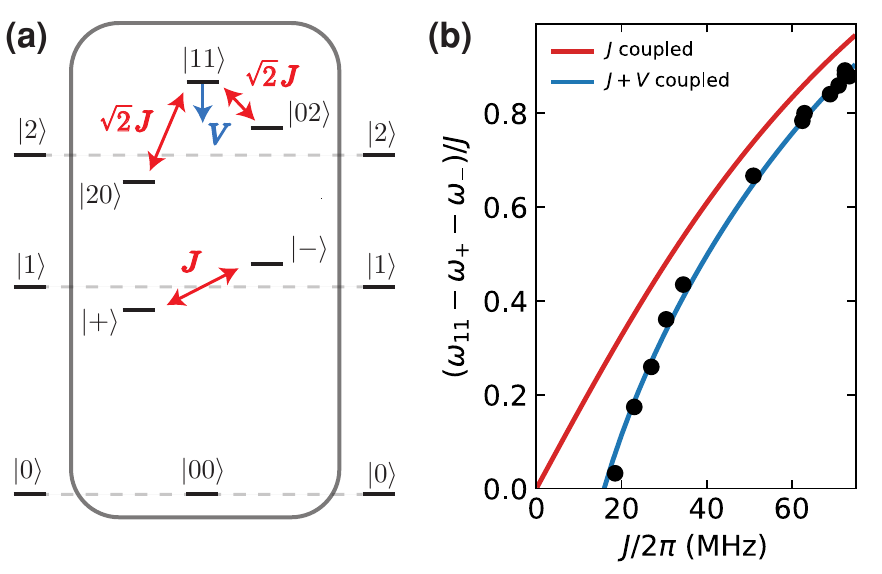}
  \end{center}
  \caption{
  {\bf Observation of nonlinear cross-Kerr coupling in spectroscopy.} 
{\bf (a)}~Level schematic of two coupled transmons up to the two-excitation manifold.
At the single-photon level, the linear coupling $J$ results in an avoided crossing between the dressed states $|+\rangle~=~\frac{|01\rangle+|10\rangle}{\sqrt{2}}$ and $|-\rangle~=~\frac{|01\rangle-|10\rangle}{\sqrt{2}}$.
As $J$ becomes comparable to the transmon anharmonicity, the $|02\rangle$, $|20\rangle$ and $|11\rangle$ levels mix with each other, resulting in an effective repulsion of ${|11\rangle}$.
On the other hand, a qubit-qubit interaction with negative cross-Kerr coupling results in lowering the energy of $|11\rangle$.
{\bf (b)}~Combined two-tone spectroscopy data (black dots) showing $\frac{\omega_{11}-\omega_{-}-\omega_{+} }{J}$ vs $J$.
The red curve is theory prediction assuming only hopping interaction between two transmons ($V=0$), while the blue one shows simulation results obtained by taking into account also the higher-order nonlinear contributions, $\frac{V}{4}(a+a^\dagger)^2\ (b+b^\dagger)^2$, which include the dominant cross-Kerr term.
The parameters used are listed in Table~\ref{tab:Parameters}.
}
  \label{fig:Cross-Kerr}
\end{figure}

As already discussed, a key feature of our implementation which differentiates it from previous tuneable couplers, is the nonlinear cross-Kerr interaction which can be tuned into different coupling regimes relative to the hopping strength and does not zero when $J$ does.
This cross-Kerr coupling, which in different contexts is referred to as $\sigma_z\sigma_z$, longitudinal~\cite{richer2016circuit}, or dispersive~\cite{blais2004cavity}, does not involve excitation hopping processes.
Its presence, however, does influence the dynamics as the occupation at one site can alter the energy level spectrum of a neighbouring site, in a process analogous to photon scattering~\cite{jin2013photon}.

In a coupled two-qubit system, the effect of cross-Kerr interaction can be seen as a shift of the energy level of the $|11\rangle$ state and can be determined spectroscopically from the transition energies relative to the ground state $\omega_{11}-\omega_{-}-\omega_{+} $.
For weakly anharmonic systems, such as the transmon, this picture becomes more complicated in the presence of linear hopping (Fig.~\ref{fig:Cross-Kerr}a).
A three-state analysis at the two-excitation manifold reveals that $|11\rangle$, $|02\rangle$ and $|20\rangle$ also couple to each other, resulting in an effective upwards repulsion of the $|11\rangle$ state~\cite{strauch2003quantum}, which scales as $\sim J^2/E_\text{C}$~\cite{geller2015tunable}.
Because the direction of this effect competes with the negative cross-Kerr shift, when the effects are similar in size, it can hinder the observation of cross-Kerr coupling in an individual spectroscopy measurement.
To separate the two effects, it is therefore necessary to measure the shift for different coupling levels.

We experimentally demonstrate the presence of cross-Kerr interaction between the two transmons, by measuring transitions in the two-excitation manifold of the coupled system.
More specifically, we extract the frequency shift of ${|11\rangle}$ at different couplings from a series of two-tone spectroscopy measurements (see Supplementary Fig.~S7), focusing on the inductively dominated regime $0\leqslant\Phi_3/\Phi_0\leqslant0.25$.
In order to distinguish between the negative cross-Kerr shift and the positive shift from linear coupling $J$, we plot $\frac{\omega_{11}-\omega_{-}-\omega_{+} }{J}$ as a function of $J$, in Fig.~\ref{fig:Cross-Kerr}b.
The red curve is theoretical prediction assuming only hopping interaction ($V=0$) between the two transmons.
The blue curve shows numerical results after diagonalising the transmon-transmon Hamiltonian with the full nonlinear coupling terms $\frac{V}{4}(a+a^\dagger)^2\ (b+b^\dagger)^2$, which includes the dominant cross-Kerr interaction.
We use the parameters listed in the second column of Table~\ref{tab:Parameters}, which differ slightly from the fitted parameters of Fig.~\ref{fig:TunCoupling}e to accommodate the effects of extra higher-order terms (see later in Fig.~\ref{fig:fit_full}).
At $J=0$ ($\Phi_3\sim0.3\ \Phi_0$) the cross-Kerr coupling $|V|/2\pi$ is around 4~MHz, and it reaches a maximum of 10~MHz at $\Phi_3=0$.
We were unable to explore the region $J/2\pi<20$~MHz in this device, due to a higher transition of the lower frequency sloshing mode hybridising with the qubits (see Supplementary Fig.~S3).

\subsection{Qubit coherence}

\begin{figure}[t]
  \begin{center}
  \includegraphics[width=\linewidth]{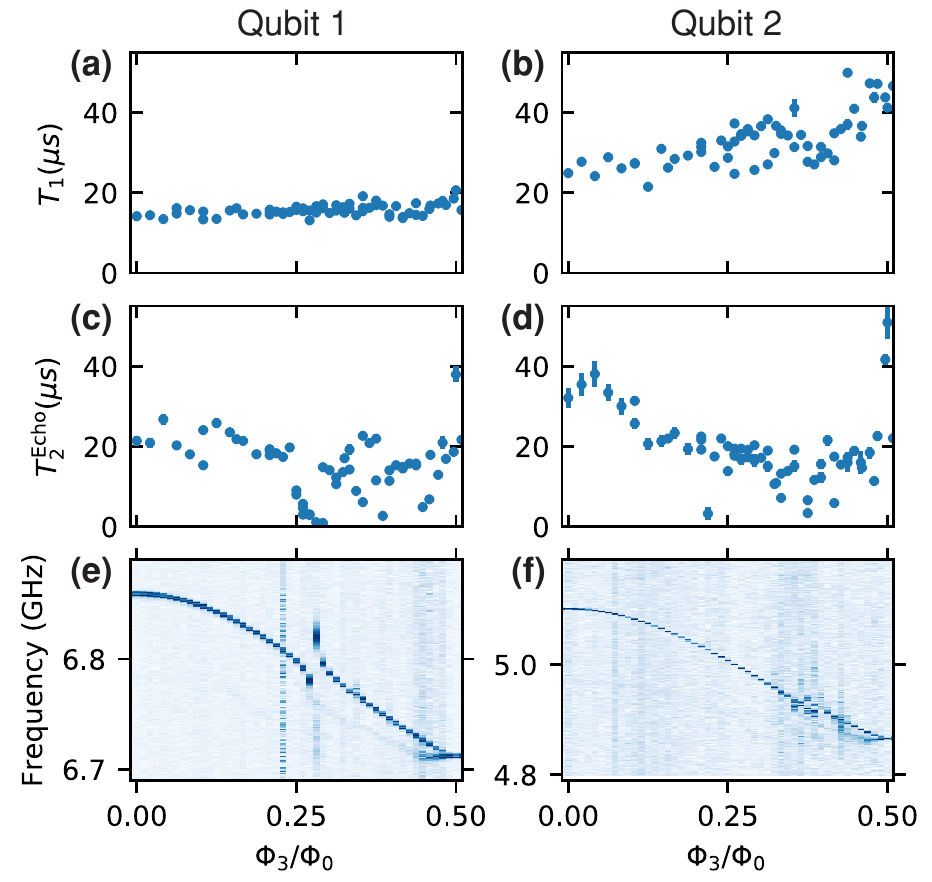}
  \end{center}
  \caption{
  {\bf Observation of high qubit coherence vs coupler flux bias.}{\bf~(a), (b)}~$T_1$ measurements showing high energy-relaxation times ($15-40\mu s$) for the whole range of coupler bias, with the qubits detuned at their top and bottom sweetspots.{\bf~(c), (d)}~Respective measurements of $T_2^{\text{Echo}}$ decay times vs $\Phi_3/\Phi_0$. High coherence is observed except for the points where a lower sloshing circuit mode crosses the qubits (see text for details).{\bf~(e), (f)}~Respective spectroscopy of both qubits vs $\Phi_3/\Phi_0$. As the inductance of the coupler is varied the qubit frequencies change as expected from theory. The 0-3 transition of the lower circuit sloshing mode crossing both qubits at $\Phi_3/\Phi_0\sim0.28$ and $\Phi_3/\Phi_0\sim0.38$ is also clearly seen. 
  }
  \label{fig:Coherence}
\end{figure}

Maintaining high coherence for all interaction strengths is an essential requirement for future implementations based on our building block device.
In Fig.~\ref{fig:Coherence}, we investigate the individual qubit properties as a function of the coupler bias, with the transmons far detuned from each other by $\sim1.8$~GHz, at their flux insensitive top and bottom sweetspots $(\Phi_1=0,~\Phi_2/\Phi_0=0.5)$.
In Figs.~\ref{fig:Coherence}a,~b, we demonstrate high relaxation times $T_1$ ($15-40~\mu s$) over the entire coupling range.
We also report a systematic study of dephasing times in our device, obtained from spin-echo measurements (Figs.~\ref{fig:Coherence}c,~d).
$T_2^\text{Echo}$ times are large overall, reaching up to $40~\mu$s, except for the points where the qubits hybridise with the lower frequency sloshing mode (at $\Phi_3/\Phi_0\sim0.28$ and $\Phi_3/\Phi_0\sim0.38$ as shown in Figs.~\ref{fig:Coherence}e,~f).
Note that the qubit frequency shifts of $\sim 200$~MHz in Figs.~\ref{fig:Coherence}e,~f are due to inherent changes to the Josephson energy of each transmon as $E_\text{J}^\text{c}$ is varied~\cite{TunCouplerSOM}. 
Repeated long Ramsey measurements were performed at $\Phi_3=0$, showing a beating pattern consistent with charge dispersion in the transmon regime~\cite{koch2007charge, riste2013millisecond}.
A measurement analysis with fits to the double sinusoidal decay pattern reveals qubit dephasing times $T_2^*$ of $10-30~\mu$s for qubit 1 and $25-40~\mu$s for qubit 2, around the flux insensitive points.

\section{Discussion}

\begin{figure}[t]
  \begin{center}
  \includegraphics[width=\linewidth]{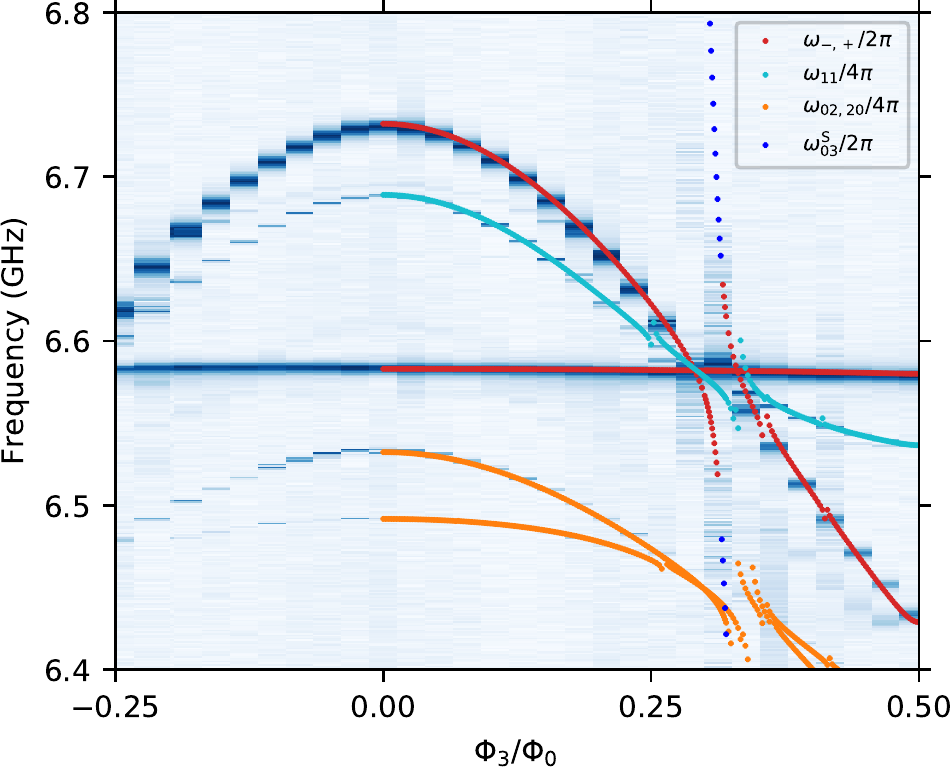}
  \end{center}
  \caption{
  { {\bf Full theoretical model of the higher excitation manifold of the device}. Measurement of the coupled system eigenspectrum (as Fig.~\ref{fig:TunCoupling}e) at higher powers. Dots show theoretical calculations using the full quantum circuit Hamiltonian including all next-to-leading order transmon-transmon coupling terms and the sloshing mode contributions, for the parameters listed in~Table~\ref{tab:Parameters}. Simulations are performed using a Hilbert space dimension of $N=15^3$.} 
  }
  \label{fig:fit_full}
\end{figure}

Our work demonstrates a key building block for circuit QED devices capable of exploring a rich vein of many-body physics in extended Hubbard models.
In the context of driven nonlinear arrays, a chemical potential term, $\mu=\omega_q-\omega_d$, could be straightforwardly implemented by coherently driving the transmons through the drive lines at a frequency $\omega_d$, which would enable the study of rich many-body phase diagrams with all $J,~V,~\mu$ tuneable~\cite{jin2013photon, jin2014steady-state}.
It may also be possible to implement topological pumping of interacting photons, by modulating the frequency of each transmon, to study bosonic transport of Fock states in a nonlinear array configuration~\cite{tangpanitanon2016topological}.
In realisations where higher-excitation manifolds might be explored, additional higher-order terms arising from the junction nonlinearity should be considered.
For example, in our implementation, the nonlinearity of the medium leads to correlated hopping terms,
\begin{align}
\label{eq:CorrelatedHopping}
\frac{V}{6} ~(\hat{a}^\dagger \hat{n}_a \hat{b}+\hat{a} \hat{n}_a\hat{b}^\dagger +\hat{b}^\dagger \hat{n}_b \hat{a}+\hat{b} \hat{n}_b\hat{a}^\dagger ),
\end{align} 
such that a photon can hop between sites, on the condition that another photon is present.
Additional contributions at higher excitation manifolds involve photon-pair tunnelling processes
\begin{align}
\label{eq:PairTunnelling}
\frac{V}{4} ~(\hat{a}^\dagger \hat{a}^\dagger  \hat{b} \hat{b} + \hat{a}\hat{a}\hat{b}^\dagger \hat{b}^\dagger),
\end{align} 
which might lead to exotic phenomena such as fractional Bloch oscillations~\cite{corrielli2013fractional}.
These contributions are explicitly derived in the Supplementary Information~\cite{TunCouplerSOM}, following a full quantum mechanical treatment of our circuit.
In Fig.~\ref{fig:fit_full}, we plot the coupled system eigenspectrum up to the two-excitation manifold, based on our full theoretical model including all next-to-leading order terms, which is found to be in excellent agreement with our data obtained at high powers.

Our circuit can also be used to study many-body effects in spin models.
When the transmon anharmonicity is much larger than the coupling strength ($E_\text{C}\gg J$), a truncation to the qubit subspace is justified, and the transmon-transmon interaction is described by a Heisenberg XXZ Hamiltonian
\begin{align}
2J(\hat{\sigma}_x\hat{\sigma}_x+\hat{\sigma}_y\hat{\sigma}_y)+V\hat{\sigma}_z\hat{\sigma}_z.
\end{align}
The coupling strengths available in this device are $J/2\pi\sim8-140$~MHz, $V/2\pi\sim0-10$~MHz, with orders of magnitude lower qubit decay rates ($3-15$~kHz).
In a slightly different design, with a larger coupling capacitor $C_\text{c}$, it would also be possible to further explore the $J\ll V$ regime (see Supplementary Fig.~S4).
One could then simulate an Ising $\hat{\sigma}_z\hat{\sigma}_z$ interaction Hamiltonian ($J=0$), with antiferromagnetic couplings of $\sim10$~MHz.
Additionally, time modulated magnetic fluxes threaded through the coupler SQUID can enable a large set of spin-spin interactions (e.g. pure $\hat{\sigma}_x\hat{\sigma}_x$ or $\hat{\sigma}_y\hat{\sigma}_y$)~\cite{sameti2017superconducting}, therefore, giving access to emulating the dynamics of almost any spin model and exploring their phase diagrams.
Connecting the coupler to four transmons on a lattice could enable simulating models with topological order such as the famous toric code~\cite{sameti2017superconducting}.
Our circuit could also be employed to engineer plaquette terms in lattice gauge theories or ring-exchange interactions in dimer models, where a longitudinal coupling much larger than the hopping term is required in order to emulate effective fields on the lattice~\cite{marcos2014two}.
Moreover, a similar architecture, featuring $\hat{\sigma}_z\hat{\sigma}_z$ coupling between transmons has been proposed theoretically for the realisation of a microwave single-photon transistor~\cite{neumeier2013single}.

In order to scale this circuit to larger lattice sizes, future experiments could take an approach where each transmon is connected to couplers via the same superconducting island, with the other island used for drive control and readout.
Using a two-island transmon design has the advantages of reducing or eliminating the number of possible current loops involving current flow across qubit junctions, as well as allowing spurious flux cross-talk to be eliminated by linear compensation techniques (see Methods).
Our coherence measurements (Fig.~\ref{fig:Coherence}) suggest that this coupler design can be realised without significantly limiting qubit coherence times, showing promise for scaling up to larger lattice sizes.

In conclusion, the implemented circuit increases the range of available interactions and phenomena that can be explored with circuit QED analog quantum simulators.
We have demonstrated hopping and cross-Kerr interactions with \textit{in situ} tuneability between two transmon qubits in a flexible and scalable superconducting circuit.
The observed decay rates are orders of magnitude lower than the coupling strengths, making this a viable platform for analog quantum simulation experiments.
Moreover, our full theoretical model of the quantum circuit is in excellent agreement with the measurements, providing a powerful tool for designing future larger scale implementations.

\section{Methods}
\subsection{Chip fabrication}
The capacitive network of superconducting islands and ground plane, together with readout and control lines are defined on a thin NbTiN film~\cite{thoen2016superconducting} on top of a high resistivity Si substrate.
The film is patterned using e-beam lithography on ARN7700 resist and etched with SF6/O2 plasma reactive-ion etching.
Josephson junctions are then fabricated on each SQUID using Al-AlOx-Al shadow evaporation following e-beam lithography patterning on a PMGI/PMMA lift-off mask and HF dip to remove surface oxides on the NbTiN contact pads.
Finally, after an e-beam patterning and reflowing PMGI step, followed by a second e-beam patterning of a MAA/PMMA resist stack, we evaporate Al airbridges which are used as cross-overs above all lines in order to ensure a uniform ground plane.

\subsection{On-chip flux cross-talk calibration}
Due to the compact geometry of our device, on-chip cross-talk between all flux channels is quite significant and extra care is required in order to decouple them.
This requirement is vital for independent control, especially for larger scale implementations where such effects could become a major experimental challenge.
We employ a systematic calibration procedure (see Supplementary~Fig.~S8) which is enabled by the fact that the frequency of the lower circuit sloshing mode (3.2~GHz at flux insensitive point) is directly associated with tuning the coupling strength via $\Phi_3$.
There is therefore one circuit degree of freedom corresponding to each bias channel.
We track spectroscopically the frequency of each degree of freedom (e.g. qubit 1) around its flux insensitive point (top sweetspot) and determine the flux offset as we vary the other two channels (2 and 3).
Repeating this for all three degrees of freedom and flux channels we were able to measure and calibrate all cross-talk effects, making all flux bias channels $\Phi_{1,2,3}$ orthogonal.
Note that the calibration method employed here allows us to distinguish between the on-chip flux cross-talk effects from the intrinsic qubit frequency shifts that are expected by varying the coupling inductance~\cite{TunCouplerSOM}.
The latter are deliberately not calibrated out in order to be able to fit the measurement data in Fig.~\ref {fig:TunCoupling}e and Fig.~\ref{fig:fit_full} with the full circuit theory Hamiltonian, however we could straightforwardly compensate for them if required.

\subsection{Device parameters}
\begin{table}[h]
\begin{tabular}{l c c}
\hline
\cline{1-3}
&  Fitting of the & Full nonlinear \\
Parameter\ \ \ \ \ \ \ \ \  & single-excitation  & circuit model\\ 
 & manifold (Fig.~\ref{fig:TunCoupling}e)  & (Figs.~\ref{fig:Cross-Kerr}b~\&~\ref{fig:fit_full})\\
\hline
\hline
$E_{\text{J}}/h$~(GHz)   & 22.99      & 23.01 \\
$C$~(fF)       & 39     & 39      \\
$C_{\text{1g}}$~(fF)     & 60.5      & 61   \\
$C_{\text{2g}}$~(fF)     & 87             & 87 \\
$E_{\text{J}}^{c(\text{max})}/h$~(GHz)   & 7.33          & 7.75  \\
$E_{\text{J}}^{c(\text{min})}/h$~(GHz)     & 0.37         & 0.39 \\
$C_{\text{c}}$~(fF)    & 18     & 20   \\
\hline
\cline{1-3}
\end{tabular}
\caption{Table of device parameters.}
 \label{tab:Parameters}
\end{table}
The device parameters are presented in Table~\ref{tab:Parameters}.
In the first column, we list the circuit parameters obtained by fitting the resonantly coupled transmon-transmon spectrum in the single-excitation manifold (Fig.~\ref{fig:TunCoupling}e) with a simplified circuit model that neglects any higher-order couplings to the sloshing mode.
The parameters in the second column are used in our full numerical model that includes all next-to-leading order terms in the circuit Hamiltonian (Supplementary Eq.~(S11)), to describe the obtained data at higher excitation manifolds (Fig.~\ref{fig:fit_full}).

\section{Data Availability Statement}
The experimental data and numerical code are available by the corresponding author upon reasonable request.

\section{Acknowledgements}
We thank D.~J.~Thoen for providing the NbTiN film, F.~Luthi and R.~Sagastizabal for experimental assistance, and Y.~Blanter, L.~DiCarlo, M.~F.~Gely and M.~J.~Hartmann for discussions.
The experiments were performed in the group of L.~DiCarlo at QuTech.
This research was supported by the Dutch Foundation for Scientific Research (NWO) through the Casimir Research School, the Frontiers of Nanoscience program, and the European Research Council (ERC) Synergy grant QC-lab.
N.K.L. also acknowledges support from the Australian Research Council through its Future Fellowship Scheme.

\section{Competing Interests}
The authors declare no conflict of interest.

\section{Author Contributions}
M.K. designed and fabricated the device with input from A.B and C.D.; M.K. performed the measurements with contributions from C.D. and N.K.L.; M.K. developed the theoretical model and analysed the data with supervision from G.A.S.; M.K. wrote the manuscript with input from all co-authors.

\end{document}


\title{Tuneable hopping and nonlinear cross-Kerr interactions in a high-coherence superconducting circuit: Supplementary information}

\author{M.~Kounalakis} 
\affiliation{Kavli Institute of Nanoscience, Delft University of Technology, Lorentzweg 1, 2628 CJ Delft, The Netherlands}
 \affiliation{QuTech, Delft University of Technology, Delft, The Netherlands}
\author{C.~Dickel}
\affiliation{QuTech, Delft University of Technology, Delft, The Netherlands}
\affiliation{Kavli Institute of Nanoscience, Delft University of Technology, Lorentzweg 1, 2628 CJ Delft, The Netherlands}
\author{A.~Bruno}
\affiliation{QuTech, Delft University of Technology, Delft, The Netherlands}
\affiliation{Kavli Institute of Nanoscience, Delft University of Technology, Lorentzweg 1, 2628 CJ Delft, The Netherlands}
\author{N.~K.~Langford}
\affiliation{QuTech, Delft University of Technology, Delft, The Netherlands}
\affiliation{Kavli Institute of Nanoscience, Delft University of Technology, Lorentzweg 1, 2628 CJ Delft, The Netherlands}
\affiliation{School of Mathematical and Physical Sciences, University of Technology Sydney, Ultimo, New South Wales 2007, Australia}
\author{G.~A.~Steele}
\affiliation{Kavli Institute of Nanoscience, Delft University of Technology, Lorentzweg 1, 2628 CJ Delft, The Netherlands}

\maketitle

\renewcommand{\theequation}{S\arabic{equation}}
\renewcommand{\thefigure}{S\arabic{figure}}
\renewcommand{\thetable}{S\arabic{table}}
\renewcommand{\bibnumfmt}[1]{[S#1]}
\renewcommand{\citenumfont}[1]{S#1}

\section{Derivation of full circuit Hamiltonian}
Here, we analytically derive the full Hamiltonian that describes the two-transmon coupled system including all circuit degrees of freedom.
We begin with a Lagrangian description of the circuit in the harmonic limit and proceed with the Hamiltonian formulation and canonical quantisation, following the methodology described in Ref.~\cite{yurke1984quantum, devoret1997edited}.

\subsection{Analytical description in the harmonic limit}

\begin{figure}[h]
  \begin{center}
  \includegraphics[width=0.5\linewidth]{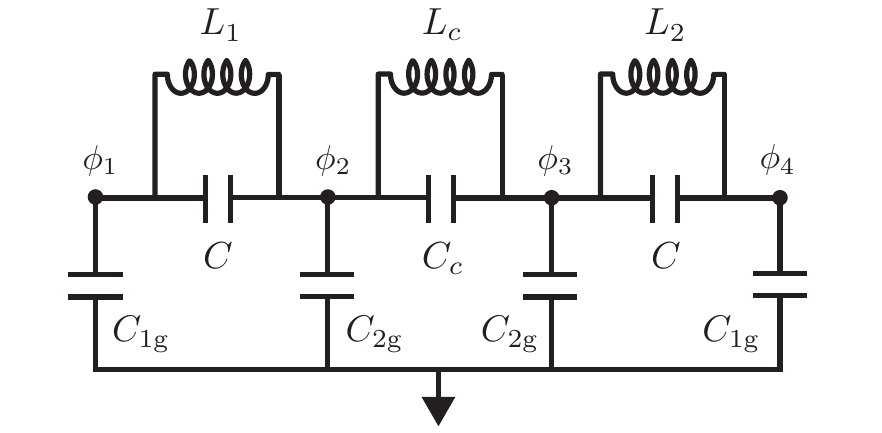}
  \end{center}
  \caption{
  {Harmonic limit of the circuit diagram shown in Fig.~1, where each Josephson element is approximated by a linear inductor.} 
  }
  \label{fig:LinearCircuit}
\end{figure}

An intuitive picture about the circuit and the relevant degrees of freedom can be drawn in the harmonic limit, where all Josephson elements are approximated by linear inductors, $L_i=\left(\frac{\phi_0}{2\pi}\right)^2\frac{1}{E_\text{J}^i}$, as shown in Fig.~\ref{fig:LinearCircuit}.
The charging and inductive energies of the linearised system are
\begin{align}
\mathcal{E}_{\text{charge}}~=~\frac{C}{2}(\dot{\phi}_2-\dot{\phi}_1)^2+\frac{C_\text{c}}{2}(\dot{\phi}_3-\dot{\phi}_2)^2+\frac{C}{2}(\dot{\phi}_4-\dot{\phi}_3)^2 + \frac{C_{1\text{g}}}{2}(\dot{\phi}_1^2+\dot{\phi}_4^2)+
\frac{C_{2\text{g}}}{2}(\dot{\phi}_2^2+\dot{\phi}_3^2),
\end{align}
and
\begin{align}
\displaystyle \mathcal{E}_{\text{ind}}~=~\frac{\left({\phi_1}-{\phi_2}\right)^2}{2L_1}+\frac{\left({\phi_3}-{\phi_4}\right)^2}{2L_2}+\frac{\left({\phi_2}-{\phi_3}\right)^2}{2L_c},
\end{align}
where the node flux $\phi_i$ (expressed in units of the reduced flux quantum $\phi_0=\hbar/2e$) is related to the potential at node $i$ as $\dot{\phi}_i=V_i$. 
The system Lagrangian, therefore, is
\begin{align}
\label{eq:L}
\mathcal{L}~=~\mathcal{E}_{\text{charge}}- \mathcal{E}_{\text{ind}}~=~\frac{1}{2}\dot{\bm{\phi}}^T[{\bf C}] \dot{\bm{\phi}}-\frac{1}{2}\bm{{\phi}}^T[{\bf L^{-1}}] \bm{{\phi}},
\end{align}
where 
\begin{align}
[{\bf C}]=
\begin{bmatrix}
   C+C_{1\text{g}}  & -C & 0 & 0 \\
    -C &C+C_{2\text{g}}+C_\text{c}  & -C_\text{c}  & 0  \\
   0 & -C_\text{c}  & C+C_{2\text{g}}+C_\text{c}   & -C \\
   0 & 0 & -C & C+C_{1\text{g}}  \\
\end{bmatrix},
\end{align}
is the capacitance matrix, and 
\begin{align}
[{\bf L^{-1}}]=
\begin{bmatrix}
   1/L_1 & -1/L_1 & 0 & 0 \\
   -1/L_1 & 1/L_1+1/L_c &-1/L_c & 0 \\
    0& -1/L_c & 1/L_2 + 1/L_c & -1/L_2  \\
   0 & 0 & -1/L_2 & 1/L_2  \\
\end{bmatrix},
\end{align}
is the inverse of the inductance matrix of the circuit, expressed in the node flux basis $\bm{\phi}^T~\dot{=}~[\phi_{1},~\phi_{2},~\phi_{3},~\phi_{4}]$.

\begin{figure}[h]
  \begin{center}
  \includegraphics[width=0.5\linewidth]{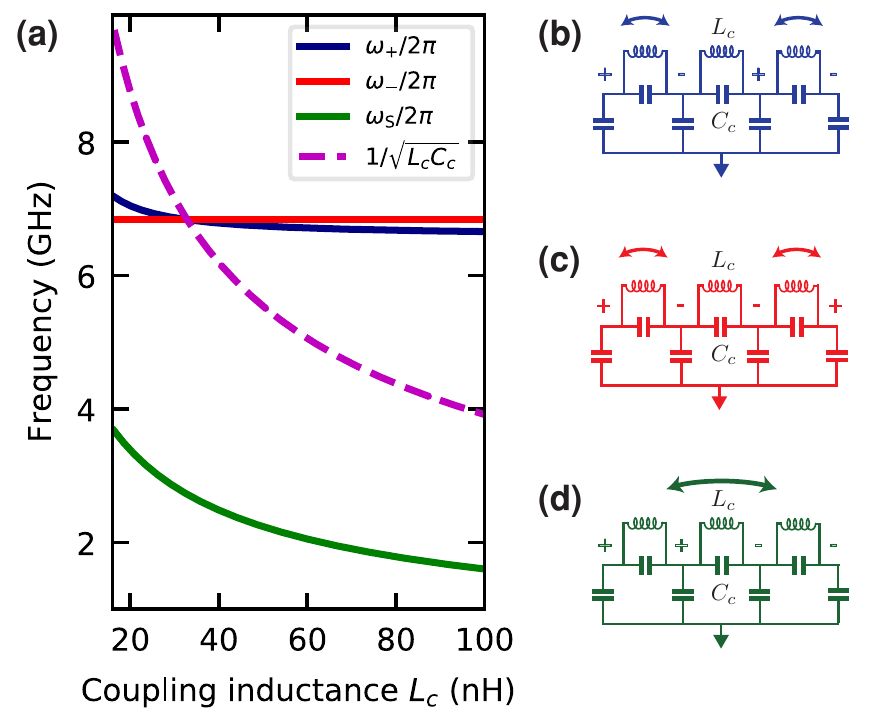}
  \end{center}
  \caption{
  {{Normal modes of the resonantly coupled transmon-transmon system in the harmonic limit.}{\bf~(a)}~Normal-mode frequencies as a function of the coupling inductance. The normal-mode splitting (blue and red curves) is suppressed at the point where the filter frequency (dashed curve) is resonant with both transmons. The sloshing mode frequency (green curve) vanishes at the limit of no inductive coupling ($L_c\rightarrow\infty$). {\bf~(b) - (d)}~Schematic representation of the circuit charge oscillations associated with each normal mode. Note that the antisymmetric mode is associated with charge oscillations across the transmon inductors only and therefore its frequency $\omega_-/2\pi$ is independent of the coupling inductance $L_c$ as shown in {\bf~(a)}.} 
  }
  \label{fig:Normal_modes}
\end{figure}

Solving the characteristic/secular equation in the resonantly coupled case ($L_1=L_2$)
\begin{align}
\label{eq:Secular}
[{\bf L^{-1}}]-\Omega^2[{\bf C}]=0,
\end{align}
yields the following normal modes (up to normalisation factors)
\begin{align}
\begin{split}
&\phi_1-a~\phi_2-a~\phi_3+\phi_4,\\
&\phi_1-b~\phi_2+b~\phi_3-\phi_4,\\
&\phi_1+c~\phi_2-c~\phi_3-\phi_4,\\
&\phi_1+d~\phi_2+d~\phi_3+\phi_4
\end{split}
\end{align}
where the coefficients $a,~b,~c,~d$ depend on the geometry of the circuit.

The first two normal modes correspond to symmetric and antisymmetric combinations of the coupled transmons as schematically depicted in Figs.~\ref{fig:Normal_modes}b,~c.
The third term corresponds to a sloshing mode associated with charge oscillations across the coupling element (Fig.~\ref{fig:Normal_modes}d).
The last term describes a zero-frequency rigid mode, corresponding to all capacitors charging up simultaneously.
The normal-mode frequencies of the linearised circuit are plotted in Fig.~\ref{fig:Normal_modes}a as a function of the coupling inductance $L_c$, for our device parameters.
Note that the normal-mode splitting and, subsequently, the coupling between the two transmons is suppressed at the point where they are both on resonance with the filter frequency $1/\sqrt{L_cC_c}$ (dashed curve in Fig.~\ref{fig:Normal_modes}a).

\subsection{Hamiltonian description of the nonlinear circuit}
The inductive energy of the nonlinear circuit (Fig.1c in the main text) is
\begin{align}
\begin{split}
\displaystyle \mathcal{E}_{\text{ind}}~=~-&E^{(1)}_{\text{J}}\cos{\left({\phi_1}-{\phi_2}\right)}-E^{(2)}_{\text{J}}\cos{\left({\phi_3}-{\phi_4}\right)}-E_{\text{J}}^{\text{c}}\cos{\left({\phi_2}-{\phi_3}\right)}\\
=~-&E^{(1)}_{\text{J}}\cos{{\psi}_{A}}-E^{(2)}_{\text{J}}\cos{{\psi}_{B}}-E_{\text{J}}^{\text{c}}\cos{\left(\frac{{\psi}_{A}-{\psi}_{B}}{2}-{\psi}_{S}\right)},
\end{split}
\end{align}
where, in the second step, we have expressed it using the transmon and sloshing mode variables $\psi_{A}~\dot{=}~\phi_{1}-\phi_{2}$, $\psi_{B}~\dot{=}~\phi_{4}-\phi_{3}$ and $\psi_{S}~\dot{=}~\frac{1}{2}\left(\phi_1+\phi_2-\phi_3-\phi_4\right)$.
The first two terms describe the Josephson energy of each transmon, while the third one describes the inductive coupling between them, as well as the inductive energy of the sloshing mode.
Here, $\psi_{A,B}$ correspond to the uncoupled transmon modes in the limit $C_\text{c}=0,~L_\text{c}\rightarrow\infty$.

Defining $\psi_{R}~\dot{=}~\frac{1}{2}\left(\phi_1+\phi_2+\phi_3+\phi_4\right)$ and performing a change of basis from $\bm{\phi}^T$ to $\bm{\psi}^T=[\psi_{R},~\psi_{S},~\psi_{B},~\psi_{A}]$, the capacitance matrix transforms as
\begin{widetext}
\begin{align}
[{\bf C^\prime}]~=~
\begin{bmatrix}
   \frac{C_{1\text{g}} + C_{2\text{g}}}{8} & 0 & \frac{C_{1\text{g}} - C_{2\text{g}}}{8} & \frac{C_{1\text{g}} - C_{2\text{g}}}{8}\\
   0 &\frac{C_{1\text{g}} + C_{2\text{g}} + 2 C_{\text{c}}}{8} &\frac{-C_{1\text{g}} + C_{2\text{g}} + 2 C_\text{c}}{8} & \frac{C_{1\text{g}} - C_{2\text{g}} - 2 C_\text{c}}{8}\\
   \frac{C_{1\text{g}} - C_{2\text{g}}}{8} & \frac{-C_{1\text{g}} + C_{2\text{g}} + 2 C_\text{c}}{8} & C +        \frac{C_{1\text{g}} + C_{2\text{g}} + C_\text{c}}{4} & \frac{-C_\text{c}}{4} \\
   \frac{C_{1\text{g}} - C_{2\text{g}}}{8} &\frac{C_{1\text{g}} - C_{2\text{g}} - 2 C_\text{c}}{8} &\frac{-C_\text{c}}{4} & C + \frac{C_{1\text{g}} + C_{2\text{g}} + C_\text{c})}{4}\\
\end{bmatrix},
\end{align}
\end{widetext}
Therefore, the system Lagrangian, expressed in the mode variable basis $\bm{\psi}$, is
\begin{align}
\mathcal{L} =&\frac{1}{2}\dot{\bm{\psi}}^T[{\bf C^\prime}]\dot{\bm{\psi}}+E^{(1)}_{\text{J}}\cos{{\psi}_{A}}-E^{(2)}_{\text{J}}\cos{{\psi}_{B}}-E_{\text{J}}^{\text{c}}\cos{\left(\frac{{\psi}_{A}-{\psi}_{B}}{2}-{\psi}_{S}\right)}.
\end{align}

The conjugate momenta $Q_{i}=\frac{\partial \mathcal{L}}{\partial \dot{\psi}_{i}}$, describing the charges associated with each mode, can be determined by inverting the capacitance matrix, since $\bm{Q}= [{\bf C^\prime}]\dot{\bm{\psi}}$.
Performing a Legendre transformation $\mathcal{H}=\sum_i\dot{\psi}_i Q_i-\mathcal{L}$ and promoting all variables to operators, satisfying $[\hat{\psi}_{i}, \hat{Q}_{i}]=i\hbar$, we obtain the full circuit Hamiltonian:
\begin{widetext}
\begin{align}
\label{eq:Hfull}
\nonumber
\hat{\mathcal{H}} &=\frac{\hat{Q}_{R}^2}{2\widetilde{C}_{R}}+\frac{\hat{Q}_{S}^2}{2\widetilde{C}_{S}}+\frac{\hat{Q}_{A}^2}{2\widetilde{C}}+\frac{\hat{Q}_{B}^2}{2\widetilde{C}}+\frac{C_\text{c}C_\text{1g}^2}{4\text{Det}[{\bf C^\prime}]} \hat{Q}_{A} \hat{Q}_{B}
+\frac{1}{\widetilde{C}_{ABS}}~\hat{Q}_\text{S}(\hat{Q}_{A}-\hat{Q}_{B})+\frac{1}{\widetilde{C}_{ABR}}~\hat{Q}_\text{R}(\hat{Q}_{A}+\hat{Q}_{B})\\
&\quad-E^{(1)}_{\text{J}}\cos{\hat{\psi}_{A}}-E^{(2)}_{\text{J}}\cos{\hat{\psi}_{B}}-E_{\text{J}}^{\text{c}}\cos{\left(\frac{\hat{\psi}_{A}-\hat{\psi}_{B}}{2}-\hat{\psi}_{S}\right)},
\end{align}
where
\begin{align}
&\widetilde{C}~=~\frac{\text{4Det}[\mathcal{C}^\prime]}{(C_\text{1g}C_\text{2g}(C_\text{1g} + C_\text{2g}) + C_\text{1g} (C_\text{1g} +  2 C_\text{2g}) C_\text{c} + C (C_\text{1g} + C_\text{2g}) (C_\text{1g} + C_\text{2g} + 2 C_\text{c}))},\\
&\widetilde{C}_{\text{S}}~=~\frac{2(C_\text{1g} (C_\text{2g} + 2 C_\text{c}) + C (C_\text{1g} + C_\text{2g} + 2 C_\text{c}))}{(4 C + C_\text{1g} + C_\text{2g} + 2 C_\text{c})}, \\
&\widetilde{C}_{\text{R}}~=~\frac{2C_\text{1g} C_\text{2g}  + 2C (C_\text{1g} + C_\text{2g})}{(4 C + C_\text{1g} + C_\text{2g} )}, \\
&{\widetilde{C}_{ABS}}~=~\frac{ 2 (C_\text{1g} ( C_\text{2g}+2 C_\text{c} ) + C (C_\text{1g} + C_\text{2g}+2 C_\text{c} ) )}{(C_\text{2g} - C_\text{1g} + 2 C_\text{c})},\\
&{\widetilde{C}_{ABR}}~=~\frac{2 C\text{1g}  C\text{2g}  + 2 C (C\text{1g}  + C\text{2g} )}{C\text{2g}  - C\text{1g} },
\end{align}
and
\begin{align}
\text{Det}[{\bf C^\prime}]&=\frac{(C_\text{1g} C_\text{2g} + C (C_\text{1g} + C_\text{2g})) (C_\text{1g} (C_\text{2g} + 2 C_\text{c}) + C (C_\text{1g} + C_\text{2g} + 2 C_\text{c}))}{4}
\end{align}
is the determinant of the capacitance matrix. 
\end{widetext}

\subsubsection{Inductive coupling contributions}
The linear contributions to the inductive coupling, arising from the first-order expansion of $E_{\text{J}}^{\text{c}}{\left(\frac{{\psi}_{A}-{\psi}_{B}}{2}-{\psi}_{S}\right)}$, are,
\begin{align}
\frac{E_{\text{J}}^{\text{c}}}{2!}{\left(\frac{\hat{\psi}_{A}-\hat{\psi}_{B}}{2}-\hat{\psi}_{S}\right)}^2=-\frac{E_{\text{J}}^{\text{c}}}{4}\hat{\psi}_{A}\hat{\psi}_{B}-\frac{E_{\text{J}}^{\text{c}}}{2}(\hat{\psi}_{A}-\hat{\psi}_{B})\hat{\psi}_{S}+\frac{E_{\text{J}}^{\text{c}}}{2}\left[\left(\frac{\hat{\psi}_{A}}{2}\right)^2+\left(\frac{\hat{\psi}_{B}}{2}\right)^2+\hat{\psi}^2_{S}\right],
\end{align}
where the first term corresponds to a direct dipole-dipole interaction between the transmons due to the coupling inductor, and the second results in an indirect transmon-transmon coupling via the off-resonant sloshing mode. The last terms describe a linear contribution to the inductive energy of each mode.

\begin{figure}[h]
  \begin{center}
  \includegraphics[width=0.7\linewidth]{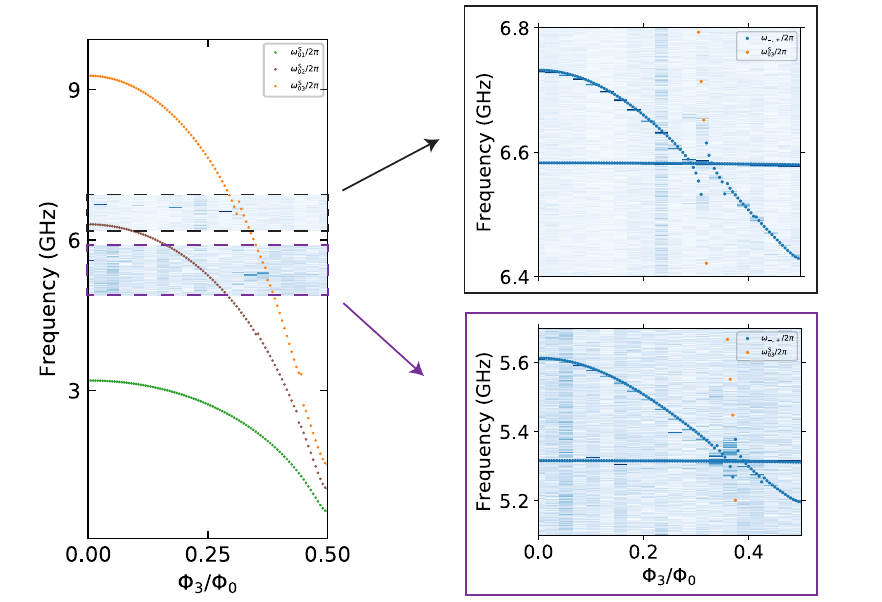}
  \end{center}
  \caption{
  {The left plot shows the expected frequency dependence of the sloshing mode and its higher levels on the coupler bias $\Phi_3/\Phi_0$.
On the right we show data and theory fit of the coupled qubit spectrum in the single-excitation manifold for two different frequency operating points (at the top and bottom sweetspots of qubit 2 and 1, respectively).
The higher $0-3$ sloshing mode transition always crosses with the qubits around the zero coupling region, for our particular choice of circuit parameters.
Note that the range of available coupling strengths is larger when operating at $\sim5.4$~GHz.} 
  }
  \label{fig:FullTheory_combo}
\end{figure}

\begin{figure}[h]
  \begin{center}
  \includegraphics[width=0.5\linewidth]{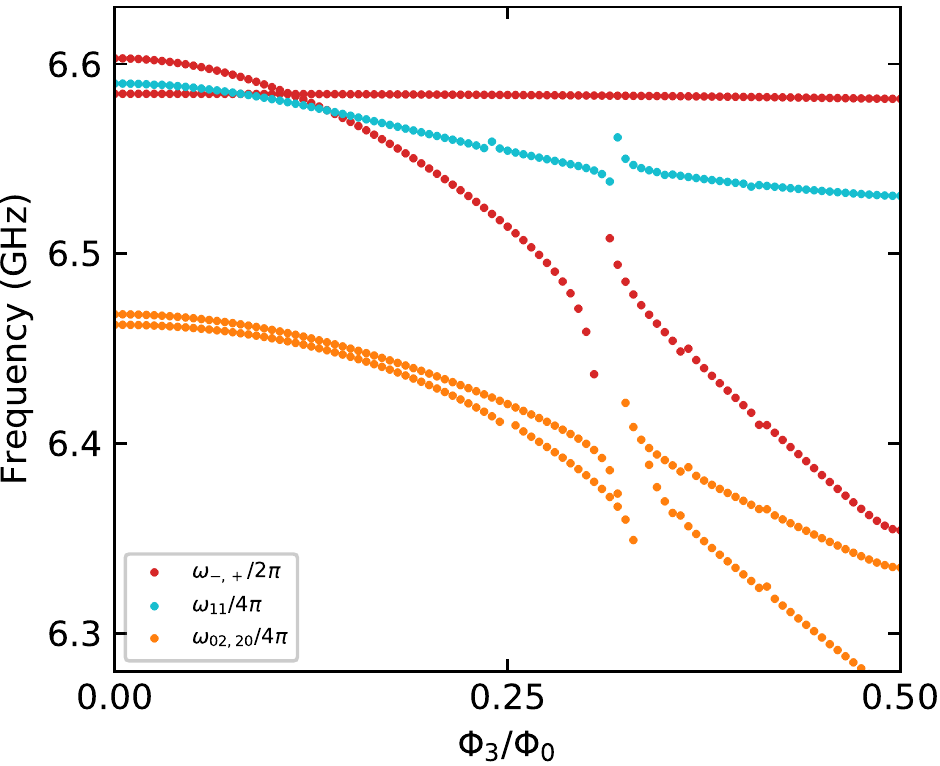}
  \end{center}
  \caption{
  {{Controlling the higher sloshing mode level crossings with an alternative circuit design}. Theoretical calculations of the coupled system energy level spectrum, demonstrate that using a slightly bigger coupling capacitor ($C_\text{c}=30$~fF), would enable experimental accessibility to regimes where $V\gg J$ and $J=0$ without other transitions interfering.} 
  }
  \label{fig:CompareCc}
\end{figure}

The second-order nonlinear contributions ($\mathcal{O}[\psi^4]$) are 
\begin{align}
\label{eq:2ndOrderInductive}
\begin{split}
\frac{E_{\text{J}}^{\text{c}}}{4!}{\left(\frac{\hat{\psi}_{A}-\hat{\psi}_{B}}{2}-\hat{\psi}_{S}\right)}^4~=~&\frac{E_{\text{J}}^{\text{c}}}{96}(\hat{\psi}^3_{A}\hat{\psi}_{B}+\hat{\psi}^3_{B}\hat{\psi}_{A})-\frac{E_{\text{J}}^{\text{c}}}{64}\hat{\psi}^2_{A}\hat{\psi}^2_{B}\\
&-\frac{E_{\text{J}}^{\text{c}}}{16}(\hat{\psi}_{A}-\hat{\psi}_{B})^2\hat{\psi}^2_{S}+\frac{E_{\text{J}}^{\text{c}}}{12}(\hat{\psi}_{A}-\hat{\psi}_{B})\hat{\psi}^3_{S}\\
&\quad\frac{E_{\text{J}}^{\text{c}}}{4!}\left[\left(\frac{\hat{\psi}_{A}}{2}\right)^4+\left(\frac{\hat{\psi}_{B}}{2}\right)^4+\hat{\psi}^4_{S}\right],
\end{split}
\end{align}
where the first direct transmon-transmon coupling term describes correlated hopping processes including small corrections to the linear coupling, and the second term contains the cross-Kerr contribution.
In our desing, the sloshing mode is at much lower frequency than the transmons, therefore the third term is not directly relevant, however, the fourth term results in qubit coupling to the $0-3$ transition of the sloshing mode, which was observed in spectroscopy near the zero coupling region.
As we show in Fig.~\ref{fig:FullTheory_combo}, this transition always crosses around the zero coupling region, even when operating at different qubit frequencies.
This is an undesired effect which arises due to the fact that the two-island transmon design introduces four nodes in our circuit.
A possible solution to this, would be to carefully engineer where these higher transitions occur and make sure that they do not cross with the qubits at the critical operation points.
For example, as we show in Fig.~\ref{fig:CompareCc}, a zero linear coupling regime could be reached for a slightly different circuit design with a larger coupling capacitor $C_\text{c}$.

Notice that, all the self-inductive terms to infinite order $\mathcal{O}(\psi_{A}^2, \psi_{B}^2, \psi_{S}^2, \psi_{A}^4, \psi_{B}^4, \psi_{S}^4, ...)$ result in the following contribution to the Hamiltonian
\begin{align}
-E_{\text{J}}^{\text{c}}\cos{\hat{\psi}_S}-E_{\text{J}}^{\text{c}}\left(\cos\frac{\hat{\psi}_{A}}{2}+\cos\frac{\hat{\psi}_{B}}{2}\right).
\end{align}
The first term describes the Josephson energy of the sloshing mode, while the rest two terms account for a correction to the transmon Josephson energies as $E_{\text{J}}^{\text{c}}$ is tuned.
This change accounts for the measured transmon frequency shifts vs coupler bias $\Phi_3$, shown in Figs.~4e,~f of the main text.

\subsubsection{Capacitive coupling contributions and mode variable elimination}
The two transmons couple directly via their charge degrees of freedom as $\frac{C_\text{c}C_\text{1g}^2}{4\text{Det}[{\bf C^\prime}]} \hat{Q}_{A} \hat{Q}_{B}$.
Additionally they couple together indirectly via the off-resonant sloshing mode $\frac{1}{\widetilde{C}_{ABS}}~\hat{Q}_\text{S}(\hat{Q}_{A}-\hat{Q}_{B})$.

The mode variable $Q_R$ has no inductive energy associated with it and only contributes to the analytical dynamics in~(\ref{eq:Hfull}) by coupling with the transmons via $\frac{1}{\widetilde{C}_{ABR}}~\hat{Q}_\text{R}(\hat{Q}_{A}+\hat{Q}_{B})$.
Notably, this coupling would vanish if the two islands of each transmon had exactly the same capacitances to ground ($C_\text{1g}=C_\text{2g}$).
In our design there is a small asymmetry ($C_\text{1g}~\neq~C_\text{2g}$), therefore, this contribution should be taken into account for an accurate theoretical description of the circuit.
It is possible, however, to eliminate this variable, while taking into account the stray contributions to the coupling between the two transmons.
To illustrate this, we rewrite these coupling terms as
\begin{align}
\frac{1}{\widetilde{C}_{ABR}}~\hat{Q}_\text{R}(\hat{Q}_{A}+\hat{Q}_{B})~&=~\left(\frac{\hat{Q}_{R}}{\sqrt{2\widetilde{C}_{R}}}+\sqrt{\frac{\widetilde{C}_{R}}{2\widetilde{C}^2_{ABR}}}(\hat{Q}_{A}+\hat{Q}_{B})\right)^2-\frac{\widetilde{C}_{R}}{2\widetilde{C}^2_{ABR}}(\hat{Q}_{A}+\hat{Q}_{B})^2\\
&=~\frac{\hat{Q}^2_{\mathfrak{R}}}{{\widetilde{C}_{\mathfrak{R}}}}-\frac{\widetilde{C}_{R}}{2\widetilde{C}^2_{ABR}}(\hat{Q}_{A}^2+\hat{Q}_{B}^2)-\frac{\widetilde{C}_{R}}{\widetilde{C}^2_{ABR}}\hat{Q}_{A}\hat{Q}_{B}.
 \end{align}
The first term is derived following a redefinition of the rigid mode $\hat{Q}_\mathfrak{R}=f(\hat{Q}_{R},\hat{Q}_{A},\hat{Q}_{B})$.
This zero frequency mode has no contribution to the equations of motion of all the other modes and can, therefore, be eliminated.
The second term accounts for a renormalisation of the transmon charging energies, while the last term accounts for a small contribution to the direct capacitive coupling, which can be absorbed into a redefinition of $J_\text{C}$.
We have therefore eliminated one variable which reduces the computational power required for numerical modelling and makes the theoretical description of the circuit more elegant. 

\subsection{Circuit quantisation in the harmonic oscillator basis}

The Hamiltonian describing the uncoupled system is given by
\begin{align}
\mathcal{H}_{0}=4E_\text{C}\hat{N}_A^2-E^{(1)}_\text{J}\cos{\hat{\psi}_{A}}+4E_ \text{C}\hat{N}_B^2-E^{(2)}_\text{J}\cos{\hat{\psi}_{B}},
\end{align}
where $E_{\text{C}}=\frac{e^2 }{2 \widetilde{C}}$ is the charging energy of each transmon and $\hat{N}_{A,B}=\frac{\hat{Q}_{A,B}}{2e}$ is the number operator associated with Cooper-pairs tunnelling through each junction~\cite{devoret1997edited}. 

Introducing annihilation (creation) operators $\hat{a}^{(\dagger)}, \hat{b}^{(\dagger)}$ such that
\begin{align}
\begin{split}
&\hat{N}_A~=~i \left(\frac{E^{(1)}_\text{J}}{32E_{\text{C}}}\right)^{1/4}(\hat{a}^\dagger-\hat{a}),~\ \ \ \ \ \hat{\psi}_A~=~\left(\frac{2E_{\text{C}}}{E^{(1)}_\text{J}}\right)^{1/4}(\hat{a}+\hat{a}^\dagger),\\
&\hat{N}_B~=~i \left(\frac{E^{(2)}_\text{J}}{32E_{\text{C}}}\right)^{1/4}(\hat{b}^\dagger-\hat{b}),~\ \ \ \ \ \hat{\psi}_B~=~\left(\frac{2E_{\text{C}}}{E^{(2)}_\text{J}}\right)^{1/4}(\hat{b}+\hat{b}^\dagger),
\end{split}
\end{align} 
the uncoupled two-transmon Hamiltonian can be approximated as two independent Duffing oscillators~\cite{koch2007charge} 
\begin{align}
\mathcal{H}_0= \hbar\omega_1\hat{a}^\dagger\hat{a}-\frac{E_{\text{C}}}{2}\hat{a}^\dagger\hat{a}^\dagger\hat{a}\hat{a}+ \hbar\omega_2\hat{b}^\dagger\hat{b}-\frac{E_{\text{C}}}{2}\hat{b}^\dagger\hat{b}^\dagger\hat{b}\hat{b},
\end{align} 
at frequencies $\omega_{1,2}=\sqrt{8E^{(1,2)}_{\text{J}}E_{\text{C}}}-E_{\text{C}}$. 
Note that the sloshing mode can also be approximately described by the transmon Hamiltonian, however this approximation ceases to be valid as $E_{\text{J}}^{\text{c}}$ is tuned to zero. 

\subsubsection{Linear coupling}
When the two transmons are resonant ($E^{(1)}_{\text{J}}=E^{(2)}_{\text{J}}~\dot{=}~E_{\text{J}}$), the linear coupling terms $\hat{Q}_{A}\hat{Q}_{B},~\hat{\psi}_{A}\hat{\psi}_{B}$, result in a dipole-dipole interaction of the form
\begin{align}
J(\hat{a}^\dagger+a)(\hat{b}+\hat{b}^\dagger).
\end{align}
In the case where $J\ll\omega$ this is equivalent to single photon hopping $J(\hat{a}^\dagger\hat{b}+\hat{b}^\dagger\hat{a})$, following a rotating wave approximation (RWA).
Moreover, in the qubit subspace this is also equivalent to $2J(\hat{\sigma}_x^1\hat{\sigma}_x^2+\hat{\sigma}_y^1\hat{\sigma}_y^2)$, where $\sigma_{x,y}$ are Pauli operators.
The coupling strength is given by
\begin{align}
J=\frac{\sqrt{8E_{\text{J}}E_{\text{C}}}}{2}\left(\frac{e^2C_\text{1g}^2C_\text{c}}{8E_{\text{C}}\text{Det}[{\bf C^\prime}]}-\frac{E_{\text{J}}^{\text{c}}}{4E_{\text{J}}}\right).
\end{align}
As we have already discussed, small contributions to $J$ need to be considered as the two transmons additionally couple via the off-resonant sloshing mode ($(\hat{Q}_{A}-\hat{Q}_{B})\hat{Q}_{S}$ and $(\hat{\psi}_{A}-\hat{\psi}_{B})\hat{\psi}_{S}$ terms).
We include these contributions, as well as higher order corrections, by diagonalising the full circuit Hamiltonian~(\ref{eq:Hfull}) in our theoretical modelling of the experimental data.

\subsubsection{Nonlinear coupling}
The first term in~(\ref{eq:2ndOrderInductive}) describes correlated photon hopping terms,
\begin{align}
\frac{V}{6} ~(\hat{a}^\dagger \hat{n}_a \hat{b}+\hat{a} \hat{n}_a\hat{b}^\dagger +\hat{b}^\dagger \hat{n}_b \hat{a}+\hat{b} \hat{n}_b\hat{a}^\dagger ),
\end{align} 
where 
\begin{align}
V=  -\frac{E_{\text{J}}^{\text{c}}E_\text{C}}{8E_{\text{J}}}.
\end{align}
This term, also leads to a small correction to the dipole-dipole coupling $J\rightarrow J-\frac{V}{6}$ .

The second term in~(\ref{eq:2ndOrderInductive}) results in cross-Kerr coupling
\begin{align}
V\hat{a}^\dagger\hat{a}\hat{b}^\dagger\hat{b}, 
\end{align}
and photon-pair tunnelling processes
\begin{align}
\frac{V}{4} ~(\hat{a}^\dagger \hat{a}^\dagger  \hat{b} \hat{b} + \hat{a}\hat{a}\hat{b}^\dagger \hat{b}^\dagger).
\end{align} 

\newpage

\section{Measurement setup}
\begin{figure}[h]
  \begin{center}
  \includegraphics[width=0.8\linewidth]{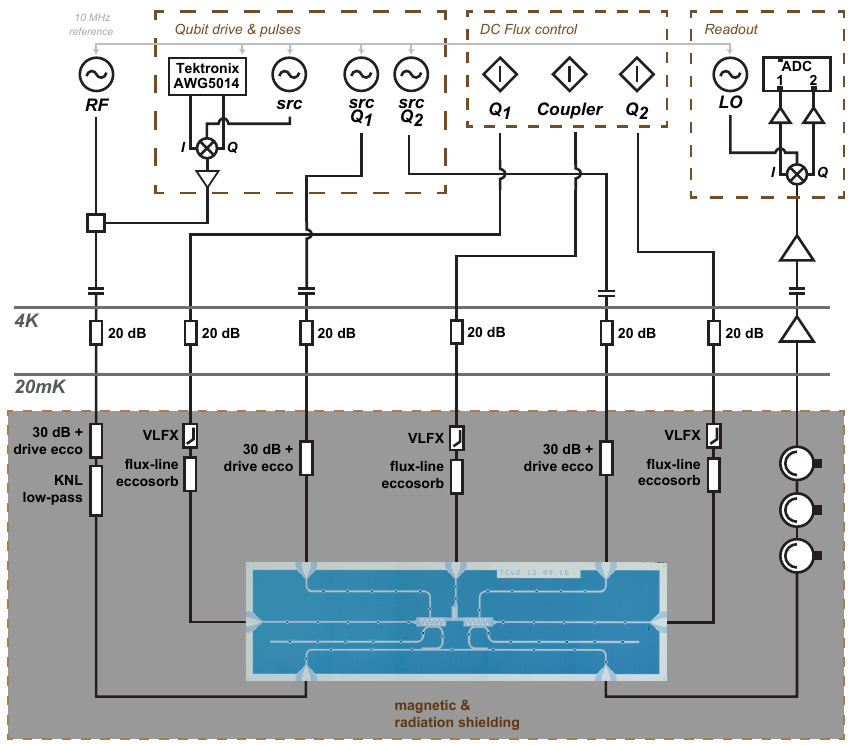}
  \end{center}
  \caption{
  Schematic of the measurement setup including microwave and DC wiring to our 2~mm~$\times$~7~mm chip.}
    \label{fig:WiringDiagram}
\end{figure}
\begin{figure}[h]
  \begin{center}
  \includegraphics[width=0.7\linewidth]{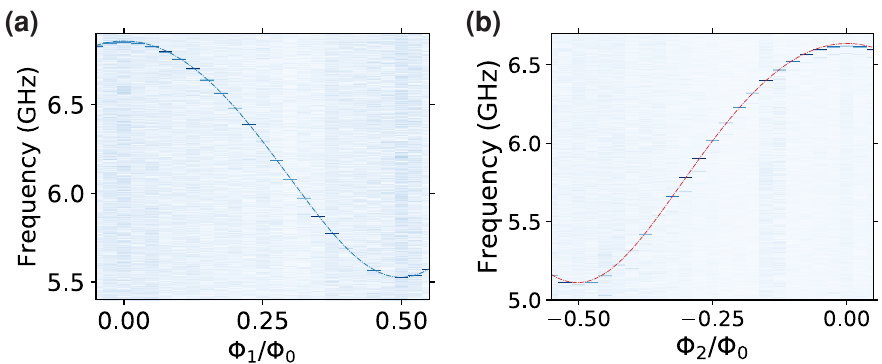}
  \end{center}
  \caption{
 Frequency tunability of {\bf (a)} qubit 1 and {\bf (b)} qubit 2 vs dedicated flux bias (raw data including theory fits). All flux channels have been made orthogonal following the calibration procedure described in Fig.~\ref{fig:FluxDecoupling}.
  }
    \label{fig:QubitArches}
\end{figure}
The measurement setup including the wiring diagram and an optical micrograph of the full 2~mm~$\times$~7~mm chip is shown in Fig.~\ref{fig:WiringDiagram}.
Our device was mounted on a printed circuit board (PCB) at the mixing chamber plate of a dilution refrigerator (fridge temperature $\sim20$~mK).
The sample was shielded against external magnetic fields, using a pair of cryogenic mu-metal shields as well as aluminium layers, and infrared radiation, using an additional copper can coated with a mixture of silicon carbide and Stycast 2850~\cite{barends2011minimizing}.
Cryogenic attenuators and home-build eccosorb filters were used in all microwave lines, resulting in a total attenuation of $\sim60$~dB for each line.
For the DC flux bias lines we use home-build eccosorb filters and commercial low-pass filters (VLFX) with 1.35~GHz frequency cutoff.
The input line additionally has a 10~GHz low-pass filter, and the output signal passes through 3 circulators, connected in series, before being amplified using cryogenic (HEMT) and room-temperature amplifiers.
The signal is then demodulated, amplified and registered with a data acquisition card.
\newpage

\section{Two-tone spectroscopy}
\begin{figure}[h]
  \begin{center}
  \includegraphics[width=\linewidth]{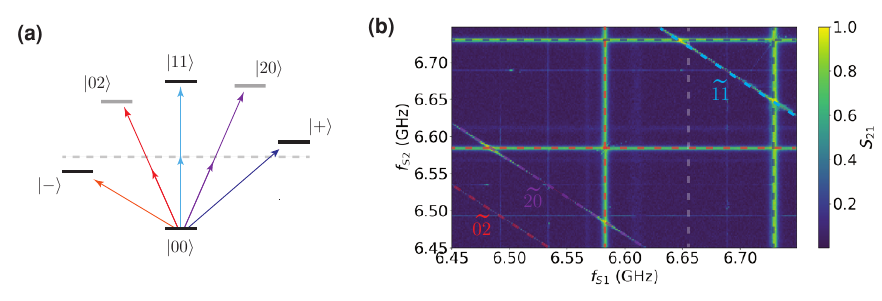}
  \end{center}
  \caption{
 {\bf (a)} Level diagram of the coupled two-transmon system on resonance. Higher level transitions are prohibited by selection rules and therefore two-tone spectroscopy is required to probe the two-exitation manifold. {\bf (b)} Two-tone spectroscopy of the coupled system at $\Phi_3/\Phi_0=0$. Vertical and horizontal lines correspond to the 01 and 10 transitions, while diagonal lines correspond to two-photon transitions.}
    \label{fig:3toneSpec}
\end{figure}

In order to determine the diagonal and cross-Kerr couplings (Fig.~3 in main text), we need to access the two-excitation manifold in the resonantly coupled system (Fig.~\ref{fig:3toneSpec}a).
However, directly exciting the even states with one source is difficult because of the transmon selection rules~\cite{BishopThesis}.
This could, in principle, be achieved via single-tone spectroscopy at very high powers (as in Fig.~5), however this results in broadening and Stark shifting of the coupled qubits frequencies, causing some uncertainty in our measurements.
We therefore perform two-tone spectroscopy, where we read out via one resonator and sweep the drive frequencies of two microwave tones ($f_\text{S1}$ and $f_\text{S2}$) applied through each driveline, as shown in Fig.~\ref{fig:3toneSpec}b.
The bright vertical lines, correspond to the dressed state transition frequencies $\omega_{+}/2\pi,~\omega_{-}/2\pi$, which can be easily excited with each source.
The diagonal $45^{\circ}$ lines correspond to two-photon transitions to $|02\rangle$, $|20\rangle$ and $|11\rangle$ as different photons from each source combine together to excite these states.
Each data point in Fig.~3b of the main text is extracted from a series of similar two-tone spectroscopy measurements.

\newpage

\section{Supplementary data}

\begin{figure}[h]
  \begin{center}
  \includegraphics[width=\linewidth]{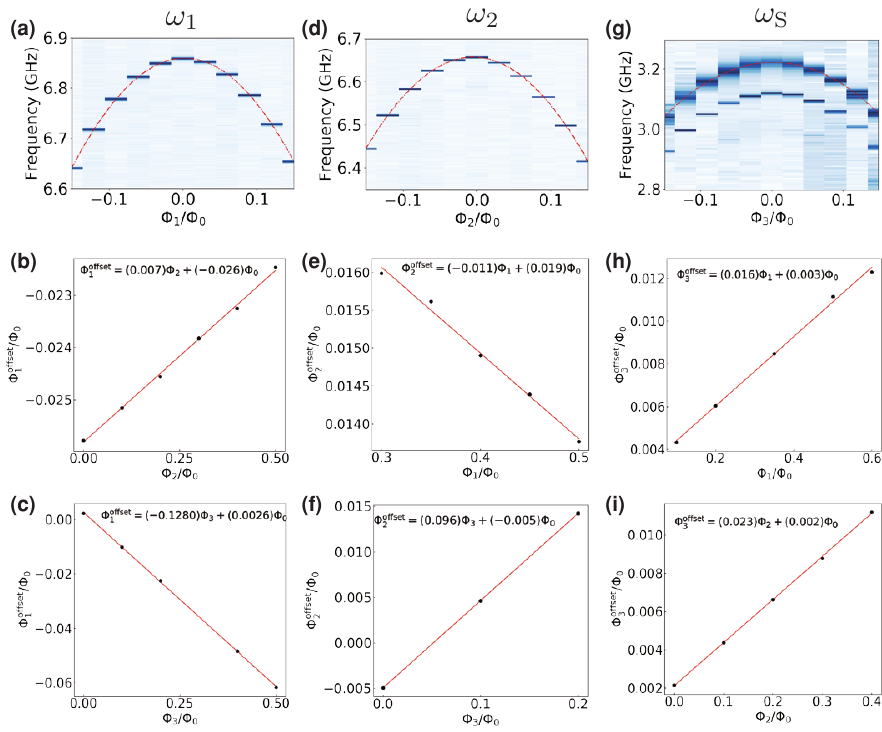}
  \end{center}
 \caption{
  {Characterisation and calibration of on-chip flux cross-talk between all bias channels.} {\bf (a)} Frequency dependence of qubit 1 vs uncalibrated flux channel 1, around its top sweetspot (for $\Phi_2/\Phi_0=0.5,~\Phi_3=0$). 
  The top sweetspot is centred at some offset value $\Phi^\text{offset}_1$, which can vary depending on the applied flux on the other channels, in the presence of on-chip cross talk.
 {\bf (b)} Dependence of $\Phi^\text{offset}_1$ on the applied flux in channel 2. Each dot is obtained from a full transmon Hamiltonian fit with $\Phi^\text{offset}_1$ as a parameter, for different $\Phi_2$ values ($\Phi_3$ is constant). A linear fit determines the dependence of $\Phi^\text{offset}_1$ on $\Phi_2$.
  {\bf (c)} Dependence of $\Phi_1^\text{offset}$ on $\Phi_3$ for constant $\Phi_2$.
  The same procedure is followed for determining $\Phi_2^\text{offset}$ vs ($\Phi_1, \Phi_3$) in {\bf (d)-(f)} and $\Phi_3^\text{offset}$ vs ($\Phi_1, \Phi_2$) in {\bf(g)-(i)}. The extracted cross-talk coefficients are then used to make all flux channels orthogonal in the experiment.
  }
  \label{fig:FluxDecoupling}
\end{figure}

\begin{figure}[h]
  \begin{center}
  \includegraphics[width=\linewidth]{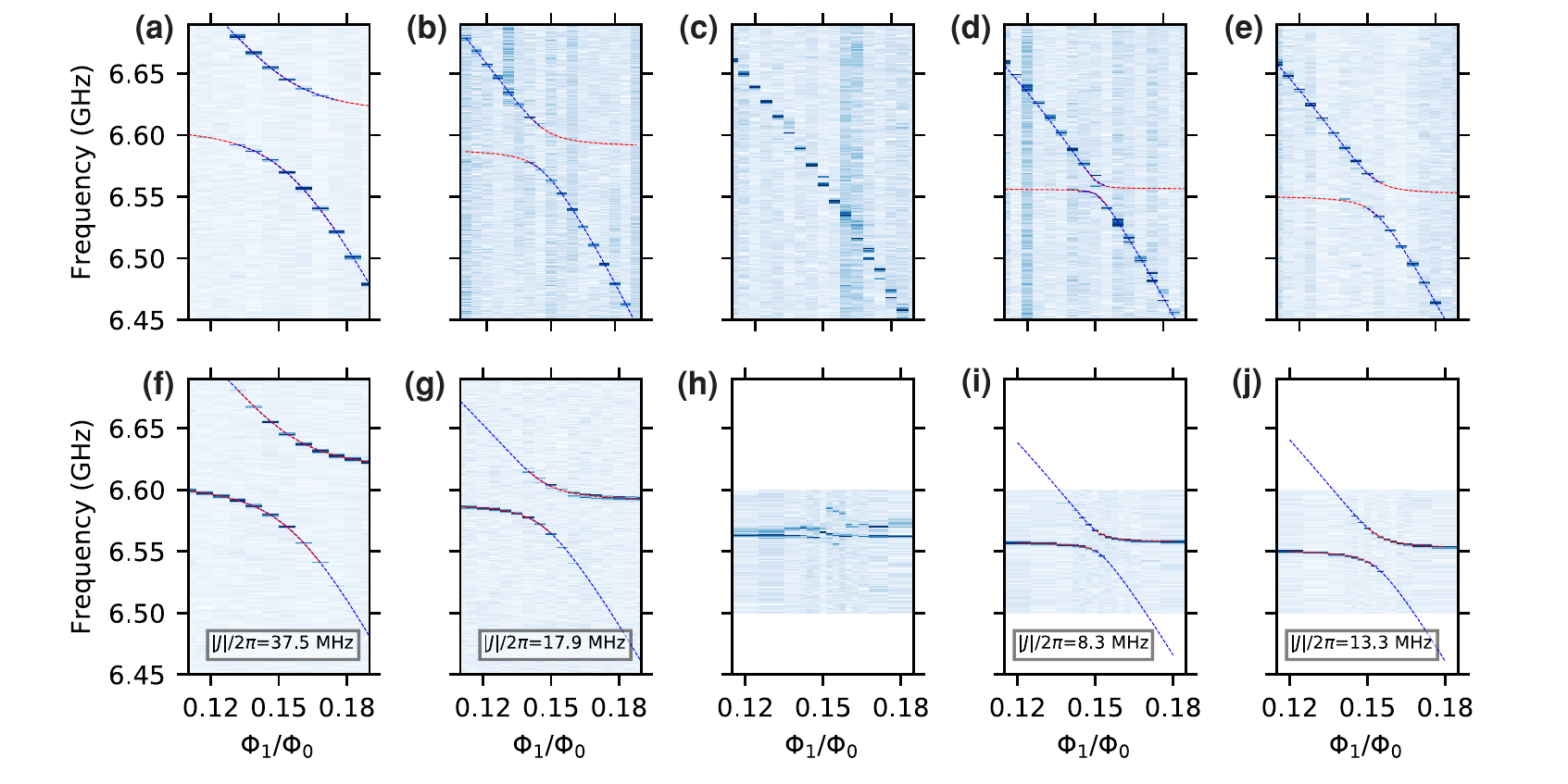}
  \end{center}
  \caption{
  {Detailed avoided crossings around the normal-mode splitting suppression point, as we tune from an inductively dominated towards a capacitively dominated coupling regime. The top row {\bf (a)-(e)} shows qubit spectroscopy data via resonator $R_1$ as qubit 1 is tuned on resonance with qubit 2, for different values of coupler bias $\Phi_3/\Phi_0~=~0.25, 0.26, 0.3, 0.32, 0.33$ (from left to right).
The bottom figures {\bf (f)-(j)} correspond to the same measurement with qubit spectroscopy via resonator $R_2$. The insets show the extracted linear coupling strengths from qubit-qubit avoided crossing fits.} 
  }
  \label{fig:DetailedAvCross}
\end{figure}